\documentclass[twocolumn]{jpsj2} 
\title{NMR measurements of intrinsic spin susceptibility in LaFeAsO$_{0.9}$F$_{0.1}$}

\author{Takashi  \textsc{Imai}$^{1,2}$\thanks{E-mail: imai@mcmaster.ca}
\thanks{Invited Paper to International Symposium on Iron-oxypnictide Superconductor, Tokyo, June 28-30, 2008},
 Kanagasingham \textsc{Ahilan}$^{1}$, Fanlong \textsc{Ning}$^{1}$, Michael A.  \textsc{McGuire}$^{3}$, Athena S.  \textsc{Sefat}$^{3}$, Ronying  \textsc{Jin}$^{3}$,  Brian C.  \textsc{Sales}$^{3}$, and David  \textsc{Mandrus}$^{3}$}

\inst{$^{1}$Department of Physics and Astronomy, McMaster University, Hamilton, Ontario L8S4M1, Canada\\
$^{2}$Canadian Institute for Advanced Research, Toronto, Ontario M5G1Z8, Canada \\
$^{3}$Materials Science and Technology Division, Oak Ridge National Laboratory, TN 37831, USA}

\abst{We will probe the intrinsic behavior of spin susceptibility $\chi_{spin}$ in the LaFeAsO$_{1-x}$F$_{x}$ superconductor ($x\sim 0.1$, $T_{c}\sim 27$~K) using $^{19}$F and $^{75}$As NMR techniques.  Our new results  firmly establish a pseudo-gap behavior with $\Delta_{PG}/k_{B} \sim 140$~K.  The estimated magnitude of $\chi_{spin}$ at 290~K, $\chi_{spin} \sim 1.8 \times 10^{-4}$~emu/mol-Fe, is approximately twice larger than that in high $T_c$ cuprates.  We also show that $\chi_{spin}$ levels off below $\sim$50~K down to $T_c$.  
}

\kword{iron pnictide superconductor, high temperature superconductivity, NMR}

\begin{document}
\maketitle

\section{Introduction}

Those who worked in the research field of copper-oxide high temperature superconductors in the 1980's are astonished by the fast pace of research into new iron-pnictide superconductors \cite{kamihara,sefat1,sm,ren,rotter,sefat2,sefat3}.  There is no doubt that proliferation of sophisticated commercial equipment around the world, such as SQUID magnetometers, is contributing to the fast progress.  However, certain aspects of condensed matter research hardly change over time.  A prime example is the difficulty in determining the intrinsic temperature dependence of spin susceptibility, $\chi_{spin}$, in transition metal composites, especially at low temperatures.  $\chi_{spin}$ is one of the most fundamental and important physical properties of solids, including strongly correlated electron systems.  Nonetheless, SQUID measurements of bulk-averaged magnetic susceptibility $\chi_{bulk}$ alone often results in misleading conclusions.  Note that $\chi_{bulk}$ is the summation of many different contributions averaged over a sample,\cite{ashcroft,obata,jaccarino}
\begin{equation}
          \chi_{bulk} = (\chi_{spin} + \chi_{orb} + \chi_{dia}) + (\chi_{defect} +  \chi_{impurity}),
\label{1}
\end{equation}
where $\chi_{orb}$ and $\chi_{dia}$ represent the paramagnetic {\it orbital} (or {\it Van Vleck}) contribution and {\it diamagnetic} contribution, respectively.  Generally, these two terms are temperature independent, but often affect interpretation of $\chi_{spin}$ when the magnitude of $\chi_{spin}$ is comparable to or less than these terms.  This is often the case in 3d transition metal composites, including high $T_c$ cuprates.\cite{pennington,mila}  $\chi_{defect}$ and $\chi_{impurity}$ represent extrinsic contributions from defect spins within the target material and impurity phases, respectively.  Separating all these different contributions is not an easy task.  In particular, $\chi_{impurity}$ could be quite large in the iron-pnictide superconductors, and a trace amount of contamination by ferromagnetic impurity phases completely ruins the SQUID data.  This explains the dearth of experimental reports on the paramagnetic properties of iron-pnictide superconductors to date. 

NMR (Nuclear Magnetic Resonance) Knight shift $K$ measurements are suited to overcoming these difficulties.   In general, one can express $K$ as \cite{jaccarino,clogston,pennington,mila}

\begin{align}
          K(T) &= K_{spin}(T) + K_{chem},\\
          K_{spin}(T) &= \sum_{i}{\frac{A_{hf}^{i}}{N_{AV}\mu_{B}} \chi_{spin}^{i}(T)},
\label{1}
\end{align}
where $K_{spin}(T)$ is the spin contribution to $K$ at temperature $T$, $N_{AV}$ is the Avogadro's number, and $K_{chem}$ is the {\it chemical shift} arising from the motion of electrons.  This term is associated with $\chi_{dia}$ and $\chi_{orb}$, and is temperature independent.  ($K_{chem}$ is sometimes represented as the Van Vleck term $K_{VV}$ or the orbital term $K_{orb}$ under certain contexts).  When we conduct NMR measurements, we apply an external magnetic field $B_{ext}$, which polarizes both nuclear and electron spins through the Zeeman interactions.  The degree of electron spin  polarization is proportional to $\chi_{spin}$.  The polarized electrons then interact with nuclear spins through hyperfine interactions. In Eq.(3), $A_{hf}^{i}$ represents the hyperfine interactions\cite{jaccarino,cohen} between observed nuclear spins ({\it e.g.} $^{19}$F nuclear spins) and surrounding electron spins ({\it e.g.} 3d electrons in FeAs layers).  {\it Since $A_{hf}^{i}$ is a temperature independent constant, one can detect the intrinsic temperature dependence of  $\chi_{spin}$ through that of $K_{spin}$}.  The superscript $i$ is attached to distinguish contributions from different sources of $\chi_{spin}^{i}$.  For example, the presence of two different types of Fermi surfaces\cite{singh} with electron and hole-like characteristics may lead to two separate contributions to $\chi_{spin}^{i}$, in analogy with the case of Sr$_2$RuO$_4$.\cite{imaiRu}  In addition, in RFeAsO$_{1-x}$F$_{x}$ (where R represents rare earth Nd, Sm, La {\it etc.}), large dipole fields from rare earth magnetic moments may also contribute to $K_{spin}$ in addition to intrinsic contributions from FeAs layers.   In the present study, we choose LaFeAsO$_{0.9}$F$_{0.1}$ to avoid contributions from magnetic moments of Nd {\it etc.}    

A major advantage of Knight shift measurements is that nuclear spins act as a {\it local} probe.  For example, for the sake of argument, let's assume that 5\% of the volume of a LaFeAsO$_{0.9}$F$_{0.1}$ sample consists of an impurity phase with a large magnetic susceptibility.  Since nuclear spins within LaFeAsO$_{0.9}$F$_{0.1}$ sample have negligible interactions with electrons in such an impurity phase, the $^{19}$F NMR Knight shift would be unaffected by the impurity spins.  This consideration alone would motivate the NMR {\it resonators} to venture into the research field of iron-pnictide superconductors, and in fact many {\it resonators} dove into the new arena.  As explained in detail below, however, accurate measurements of $K$ in polycrystalline samples turned out to be more challenging than {\it resonators} expected.

If the measurements of intrinsic $\chi_{spin}$ and $K$ are so tricky, why should we bother to measure them?  At the outbreak of research into iron-pnictide superconductors earlier this year, the first question we asked to ourselves was, {\it does LaFeAsO$_{1-x}$F$_{x}$ show evidence for electron-electron correlations?}  Given that Fe is an itinerant ferromagnet and FeAs is an itinerant antiferromagnet,\cite{feas} it would be natural to speculate on the presence of strong electron-electron correlations in superconducting LaFeAsO$_{1-x}$F$_{x}$, too; if that is indeed the case, the superconducting mechanism may be quite exotic.  For example, if ferromagnetic spin correlations are strong and electron spins tend to line up, perhaps the spin-triplet, orbital p-wave symmetry\cite{unconventional} may be favored by the Cooper pairs below $T_c$.  On the other hand, the undoped parent phase LaFeAsO shows SDW (Spin Density Wave) order with a large wave vector ${\bf q}$,\cite{delacruz} and antiferromagnetic correlations may remain in the carrier-doped superconducting phase.  In such a scenario, the spin-singlet, orbital d-wave symmetry\cite{unconventional} may be favored instead, in analogy with high $T_c$ cuprate supercondcutors.  Or, perhaps, both ferromagnetic and antiferromagnetic spin correlation effects may be present, in analogy with Sr$_2$RuO$_4$.\cite{imaiRu,imaiRunote}  In any case, $\chi_{spin}$ should provide valuable insight into the nature of electrons in the normal state above $T_c$ and clues to the superconducting mechanism.  We set out to address this important issue using $^{19}$F NMR spectroscopy.\cite{ahilan} 

In this invited paper, we will describe the basic concept as well as the standard procedures of NMR Knight shift measurements using the LaFeAsO$_{0.9}$F$_{0.1}$ superconductor ($T_{c} = 27$~K) as an example.  We will proceed in a somewhat pedagogical manner so that non-experts in NMR could gain a sense on how resonators carry out NMR measurements and interpret the Knight shift data.  Our new $^{19}$F as well as $^{75}$As Knight shift data confirm our initial report,\cite{ahilan} and firmly establish the pseudogap behavior of LaFeAsO$_{1-x}$F$_{x}$ superconductors.  The rest of the paper is organized as follows.  In section 2, we describe the basics of  NMR techniques, and explain why measurements of the $^{19}$F NMR Knight shift, $^{19}K$, can be advantageous over NMR measurements of other nuclei.  In section 3, we present the $^{19}K$ data and discuss the implications.  In section 4, we compare $^{19}K$ with the $^{75}$As NMR Knight shift,  $^{75}K$.  Section 5 is the summary and future outlook.  The measurements of the Knight shift below $T_c$ is beyond the scope of the present paper, and we refer readers to our recent $^{75}K$ measurements in single crystal BaFe$_{1.8}$Co$_{0.2}$As$_{2}$ ($T_{c}=22$~K):  In that work, we demonstrated that $^{75}K$ decreases below $T_c$ for both c- and ab-axes orientations, and thereby proved the consistency with singlet pairing scenarios.\cite{ning} 

\section{Why $^{19}$F NMR?}

In NMR measurements, we place our sample in a static external magnetic field $B_{ext}$, and apply radio frequency (r.f.) pulses to excite nuclear spins in the sample.  The r.f. frequency $f$ of the pulses is tuned with the Larmor frequency of the nuclear spins, $f \sim \gamma_n B_{ext}$, where $\gamma_n$ is the nuclear gyromagnetic ratio of the particular nuclei of the interest.  The r.f. pulses induce {\it spin echo}, which results from coherent Larmor precession of the nuclear spin ensemble in $B_{ext}$.  These spin echoes produce inductive voltage signals in the NMR coil.  The NMR signal intensity is typically of the order of microvolts, and depends on the magnitude of nuclear moments being observed.  

In principle, one can excite and observe NMR signals from $^{139}$La, $^{57}$Fe, $^{75}$As, $^{17}$O,  and $^{19}$F nuclei in LaFeAsO$_{1-x}$F$_{x}$. The key properties of these NMR active nuclei are compared in Table I.  One crucial fact to notice is that $\gamma_n$ is different for different nuclei, hence so is the resonance condition $f \sim \gamma_{n} B_{ext}$.  In other words, if we apply a constant magnetic field $B_{ext}$ to our sample and search NMR signals by scanning the frequency, $^{19}$F and $^{75}$As NMR peaks appear at different frequencies ($^{19}f~ >~^{75}f$), and therefore we can observe them separately.  Alternatively, we may fix the experimental frequency $f$, and scan $B_{ext}$ to search NMR signals;  $^{19}$F and $^{75}$As NMR signals appear separately when $B_{ext} \sim f/\gamma_{n}$ is satisfied for each of them ($^{19}B_{ext}~ <~^{75}B_{ext}$).  

Each nuclear spin interacts with surrounding electrons through electron-nucleus hyperfine interactions.\cite{cohen,jaccarino} Hyperfine interactions are highly local, hence the NMR signals from solids reflect the {\it local} electronic properties where the particular nuclear spins are located.  In other words, we can use nuclear spins at different locations in the unit cell to "look into" different parts of the electronic structure in a complicated structure of solids.

\begin{table}[t]
\caption{Properties of NMR active nuclei in LaFeAsO$_{1-x}$F$_{x}$.  The units of $\gamma_{n}/2\pi$ are MHz/Tesla.  $N_A$ represents the natural abundance. The fourth entry is the NMR intensity relative to the equal number of proton $^{1}$H nuclear spins in the same magnetic field $B_{ext}$.}
\label{t1}
\begin{tabular}{c|ccccc}
\hline
Nuclei & Spin $I$ & $\gamma_{n}/2\pi$ & $N_A$~(\%) & Intensity \\
\hline
$^{139}$La & 7/2  & 6.0144  & 99.9  & $5.9\times10^{-2}$  \\ 
$^{57}$Fe & 1/2 & 1.3758 & 2.19 & $3.4\times10^{-5}$ \\
$^{75}$As & 3/2 & 7.2919  & 100 & $2.5\times10^{-2}$ \\
$^{17}$O & 5/2 & 5.772  & $3.7 \times 10^{-2}$ & $2.9\times10^{-2}$  \\
$^{19}$F & 1/2 & 40.0541  & 100 & 0.83 \\
\end{tabular}
\end{table}

Since the natural abundance of $^{57}$Fe and $^{17}$O is very low, NMR experiments on these nucei are rather difficult unless we enrich the sample with costly isotopes (one enriched sample may cost up to several thousand dollars).  This leaves $^{139}$La, $^{75}$As, and $^{19}$F as candidates for our NMR investigation.  However, $^{19}$F has a major advantage over $^{139}$La and $^{75}$As in observing the intrinsic behavior of $\chi_{spin}$ with high accuracy and relative ease: namely, {\it the nuclear spin of $^{19}$F is $I=\frac{1}{2}$}.  

In order to understand why NMR measurements on $I=\frac{1}{2}$ nuclei are suited for an accurate determination of $\chi_{spin}$, we express the NMR frequency $f$ in a high magnetic field $B_{ext}$ as\cite{jaccarino,abragam}
\begin{equation}
          f  = f_o + \Delta f = (1+K)\gamma_{n} B_{ext} + \nu_{Q}^{(1)}(\theta, \phi)+ \Delta \nu_{Q}^{(2)}(\theta, \phi,B_{ext}).
\label{2}
\end{equation}
where $f_o = \gamma_{n} B_{ext}$ is the bare NMR frequency in the absence of hyperfine interaction between nuclear spins and the surrounding environment in a given crystal structure.  $\Delta f = K\gamma_{n} B_{ext} + \nu_{Q}^{(1)}+ \Delta\nu_{Q}^{(2)}$ represents the resonance frequency shift under the presence of hyperfine interactions, arising from the {\it Knight shift} $K$, {\it first order quadrupole contribution} $\nu_{Q}^{(1)}$, and {\it second order quadrupole contribution} $\Delta\nu_{Q}^{(2)}$.  $\theta$ and $\phi$ are polar angles between $B_{ext}$ and the main principal axis of the EFG (Electric Field Gradient) tensor at the position of the observed nuclei.  To determine the Knight shift $K$, we need to measure $\Delta f$, then subtract the quadrupole contributions.

As explained in section 4 using the case of $^{75}$As NMR as an example, the quadrupole contributions $\nu_{Q}^{(1)}$ and $\Delta\nu_{Q}^{(2)}$are often orders of magnitude larger than that of  $K$.  Accordingly, proper estimation of $K$ often requires tricky and cumbersome procedures. 
However,   {\it for $I=\frac{1}{2}$,  the nuclear quadrupole interaction is always zero}, and $\nu_{Q}^{(1)} = \Delta\nu_{Q}^{(2)} =0$.  This means that for $^{19}$F the NMR frequency shift $\Delta f$ is caused entirely by the effects of the Knight shift $K$, and we can readily invert Eq.(4) to obtain $K$,
\begin{equation}
          ^{19}K=\frac{^{19}f - ^{19}f_{o}}{^{19}f_{o}}  =\frac{^{19}\Delta f}{^{19}f_{o}} = \frac{^{19}f}{^{19}\gamma_{n}B_{ext}} - 1.
\label{1}
\end{equation}
In other words, all one needs to do is: (a) measure the $^{19}$F NMR frequency $^{19}f$ from our sample in a given external field $B_{ext}$, and (b) calibrate the magnitude of $B_{ext}$ (we use the proton NMR frequency in water for the calibration); then we obtain $^{19}K$ from Eq.(5) without elaborate analysis.   Additional major advantage of working on $I=\frac{1}{2}$ spins is that, when nuclear dipole-dipole interactions are the primary cause of the distribution of the NMR frequency $f$, the line width is very narrow.  When the line width is less than the r.f. bandwidth of our NMR spectrometer, we can obtain the NMR lineshape very accurately by taking a FFT of the spin echo signal.\cite{slichter}

\section{$^{19}K$ in LaFeAsO$_{0.9}$F$_{0.1}$}

In Fig.2, we show representative $^{19}$F NMR lineshapes in LaFeAsO$_{0.9}$F$_{0.1}$ obtained by FFT of spin echo signals.  For these measurements, we mixed a finely ground polycrystalline sample with an appropriate amount of glue with low viscocity (Stycast 1266), and cured the mixture in a magnetic field of $\sim 8$~Tesla in ambient temperature.  Since the magnetic susceptibility of LaFeAsO$_{0.9}$F$_{0.1}$ has an anisotropy, $\chi_{bulk}^{ab} > \chi_{bulk}^{c} (>0)$, polycrystals rotate in the magnetic field to minimize their energy.  The ab-plane tends to be aligned with the applied magnetic field when the mixture solidifies.  We inserted the aligned polycrystalline sample in an NMR coil inside an external magnetic field $B_{ext}$, and conducted NMR measurements.  The vertical grey arrow in Fig.2 marks the bare $^{19}$F NMR frequency $^{19}f_o$ we would observe if there were no hyperfine interactions.  The actual peak of the lineshape, $^{19}f$, is shifted to the higher frequency side.  This means that $^{19}\Delta f > 0$, and hence $^{19}K_{ab} > 0$, where the suffix $ab$ means that the Knight shift is measured along the crystal $ab$-plane.  With decreasing temperature, the peak frequency $^{19}f$ gradually approaches toward  $^{19}f_o$, but $^{19}f$ hardly changes below $\sim 50$~K.  That is, {\it $^{19}K_{ab}$ decreases with temperature, and levels off below $\sim 50$~K}.  We deduced the temperature dependence of  $^{19}K_{ab}$ from Eq.(5), and summarized the results in Fig.3.  The new results agree well with our earlier results obtained for an unaligned polycrystalline sample of LaFeAsO$_{0.89}$F$_{0.11}$\cite{ahilan} within the overall experimental uncertainties of $\sim 0.002$~\%.   In what follows, we take the sign of the hyperfine coupling as positive, $^{19}A_{hf} >0$,\cite{positiveAhf} and $^{19}K_{chem}\sim 0.045$~\%.\cite{Kchem}  $^{19}A_{hf} >0$ implies that {\it $\chi_{spin}$ decreases with temperature, and levels off below $\sim 50$~K}.  $^{19}K_{chem}\sim 0.045$~\% implies that $^{19}K_{spin,ab}$, and hence $\chi_{spin} = (N_{AV} \mu_{B}~^{19}K)/^{19}A_{hf}$, decreases by a factor $\sim 2$ from 300~K to $T_c$. 

The observed temperature dependence of $^{19}K_{ab}$ leads us to several conclusions.  First, non-interacting or weakly-interacting localized magnetic moments don't exist in LaFeAsO$_{0.9}$F$_{0.1}$.  Such localized moments would result in a Curie-Weiss behavior of the Knight shift, $^{19}K_{spin} \sim C/(T-\theta)$.  Second, we find no evidence for enhancement of ferromagnetic electron-electron correlations toward $T_c$.  Growth of ferromagnetic correlations at ${\bf q} ={\bf 0}$  with decreasing temperature would manifest itself in the growth of $\chi_{spin}$, hence $^{19}K$.  Our Knight shift data rule out such a scenario.  Third, with the aid of our earlier results of the spin-lattice relaxation rate $^{19}(1/T_1T)$ at $^{19}$F sites, we can also rule out a simple picture that antiferromagnetic spin-spin correlations are the {\it primary cause} of the decrease of  $\chi_{spin}$ with temperature.\cite{ahilan}  This point deserves additional explanation.  If short range antiferromagnetic correlations grow toward a critical point with decreasing temperature in a conventional sense, neighboring spins will try to point in opposite directions.  Then the ${\bf q} ={\bf 0}$ mode of the spin susceptibility, $\chi_{spin}$, would decrease when observed by low frequency probes such as SQUID and NMR.  Note that $\chi_{spin}$ measures the tendency of all spins to point along the same orientation uniformly (i.e. ${\bf q} ={\bf 0}$).  However, in such a conventional antiferromagnetic short range order scenario, the spectral weight of the low frequency antiferromagnetic spin fluctuations generally grows with decreasing temperature too. This means that $^{19}(1/T_1T)$ at $^{19}$F sites must {\it increase} with decreasing temperature.  On the contrary, $^{19}(1/T_{1}T)$ at $^{19}$F sites shows the same temperature dependence as $^{19}K$, and {\it decreases} with temperature.\cite{ahilan}  Therefore antiferromagnetic short range order {\it alone} can't account for the observed decrease of $^{19}K$ {\it and}  $^{19}(1/T_{1}T)$ toward $T_c$.  Antiferromagnetic correlations may be certainly growing for high frequency/energy modes, but such growth has to take place at the expense of the loss of the low frequency spectral weight of  the spin-spin correlation function, $S({\bf q},f)$, toward $T_c$ ({\it i.e. "total moment sum rule"}).  Analogous behavior was previously observed in the underdoped high $T_c$ cuprates, and is known as the {\it pseudo-gap behavior}.  We emphasize that  it was our $^{19}$F NMR measurements that first arrived at this key conclusion of the pseudogap phenomenon in iron-pnictide high $T_c$ superconductors.\cite{ahilan}   

By fitting the temperature dependence of  $^{19}K_{ab}$ to
\begin{equation}
          ^{19}K_{ab} =A + B \times exp(-\Delta_{PG}/k_{B}T),
\label{1}
\end{equation}
we may quantify the magnitude of the pseudo-gap as $\Delta_{PG}/k_{B} = 140 \pm 20$~K.  This value is very close to $\Delta_{PG}/k_{B} = 172$~K as estimated by the fit of $^{75}(1/T_{1}T)$ at the $^{75}$As sites  to an analogous formula.\cite{nakai}.  In passing, we recently found that the pseudo-gap is much greater, $\Delta_{PG}/k_{B} = 560$~K, in BaFe$_{1.8}$Co$_{0.2}$As$_2$ ($T_{c}=22$~K).\cite{ning}

The origin of the leveling of  $\chi_{spin}$ below $\sim50$~K is not understood well at this time.   As emphasized first by Nakai et al. from the leveling of their $^{75}(1/T_{1}T)$ data, the {\bf q} integral of low frequency spin fluctuations also levels off below $\sim50$~K.  In fact, our earlier data of  $^{19}(1/T_{1}T)$ at $^{19}$F sites also levels off below $\sim50$~K, although we didn't emphasize it.\cite{ahilan}  In canonical Fermi-liquid systems (such as simple Cu metal {\it etc.}), quite generally\cite{jaccarino}
\begin{align}
          K_{spin} & \propto A_{hf} N(E_{F}),\\
          1/T_{1}T & \propto (A_{hf} N(E_{F}))^{2},\\
          1/T_{1}T(K_{spin})^{2} & = constant,
\label{1}
\end{align}
where $N(E_{F})$ is the density of states at the Fermi energy.  Eq.(9) is the celebrated {\it Korringa relation}, and is often used as a criterion for establishing the canonical Fermi-liquid behavior of strongly correlated electron systems.  In the present case,  the Korringa relation certainly holds below $\sim50$~K, because both $K_{spin}$ and $1/T_{1}T$ become constant.  But it is important to realize that the temperature independence of $K_{spin}$ and $1/T_{1}T$ does not necessarily prove that FeAs layers cross over to a canonical Fermi-liquid regime below $\sim50$~K.  In Fig.4, we present the temperature dependence of resistivity $\rho$ of the same sample used for NMR measurements.  If a canonical Fermi-liquid picture is valid below $\sim 50$~K, we expect to observe another signature of Fermi-liquid, $\rho \propto T^2$.  However, our resistivity data does {\it not} satisfy the $T^2$ law below $\sim 50$~K.  Since the crossover to a $\chi_{spin}=constant$ takes place right above $T_c$, the power-law fit of $\rho$ in the narrow temperature range is dicey and somewhat inconclusive.  In a recent study of a single crystalline sample of BaFe$_{1.8}$Co$_{0.2}$As$_2$ ($T_{c}=22$~K), we also demonstrated that $K_{spin} \sim constant$ and $1/T_{1}T \sim constant$ hold in a much broader temperature range below $\sim 100$~K.\cite{ning}  In that case, we found $\rho \propto T^n$ with $n \sim 1$ in the same low temperature regime above $T_c$.\cite{sefat3} 

\section{$^{75}$As NMR lineshape and $^{75}K$}
In this section, we will provide a brief account of how the Knight shift is usually measured for $I \neq \frac{1}{2}$ quadrupolar nuclei using $^{75}$As as an example.  In Fig.5, we show a field-swept $^{75}$As NMR lineshape for all three permissible transitions: the $I_z = -\frac{1}{2}$ to $+\frac{1}{2}$ central transition near $B_{ext}\sim7.5$~Tesla, and $I_z = \pm\frac{1}{2}$ to $\pm\frac{3}{2}$ satellite transitions near 8.2 and 6.8~Tesla.  Note that the horizontal axis is inverted, so that the NMR signals with positive frequency shift $\Delta f$ appear on the right hand side.  The first order quadrupole term $\nu_{Q}^{(1)} (>> \Delta \nu_{Q}^{(2)})$ has null contribution to the central transition, but shifts the satellite transitions by a large amount,\cite{abragam}
\begin{align}
          \Delta f^{center} & = K\gamma_{n} B_{ext} + \Delta\nu_{Q}^{(2)},\\
          \Delta f^{satellite} & \sim K\gamma_{n} B_{ext} \pm \nu_{Q}^{(1)},
\label{1}
\end{align}
and $\nu_{Q}^{(1)}\sim^{75}\nu_{NQR}/2$, where $^{75}\nu_{NQR}\sim 11$~MHz is the $^{75}$As NQR frequency.\cite{grafe}  Since we measured the lineshape while sweeping $B_{ext}$, the quadrupole split between two satellite transitions should be given by $2\nu_{Q}^{(1)}/^{75}\gamma_{n} \sim 11/7.2919 \sim 1.5$~Tesla, in good agreement with the experimentally observed split $(8.2 - 6.8) \sim 1.4$~Tesla.

The inset of Fig.5 shows the central transition of the unaligned and aligned polycrystalline samples measured in the same condition.  Since the nuclear quadrupole interaction affects the central transition only through the second order perturbation term $\Delta \nu_{Q}^{(2)}$, the observed linewidth is narrower than that of satellite transitions.  Thus the central transition is better suited for our purpose of Knight shift measurements.  For the unaligned sample, the polar angle dependence of the second order term $\Delta \nu_{Q}^{(2)}(\theta, \phi, B_{ext})$ results in the broad "double-horn" lineshape.\cite{abragam}  For the aligned polycrystals, we observe a pronounced peak near the low field edge $B_{ext} \sim 7.45$~Tesla, because the main principal axis of the EFG tensor points along the crystal c-axis, hence $\theta \sim \pi/2$ for aligned crystallites.  The lower field edge corresponds to $\theta \sim \pi/2$.  The small  tail of the spectrum toward higher magnetic field values is due to crystallites that failed to align when cured in Stycast 1266.

Our task is to measure the $^{75}$As Knight shift $^{75}K_{ab}$ from accurate measurement of the central peak position near 7.45~Tesla.  In order to determine $^{75}K_{ab}$ from the central transition, we need to subtract the contribution of $\Delta \nu_{Q}^{(2)}$ in Eq.(10).  A major challenge here is that, $\Delta \nu_{Q}^{(2)}$ is much greater than $K\gamma_{n} B_{ext}$ in Eq.(10).  This means that a small error in estimating the $\Delta \nu_{Q}^{(2)}$ term leads to a large error of $^{75}K_{ab}$.  By far the most reliable method to separate the effects of $\Delta \nu_{Q}^{(2)}$ accurately is to measure the field dependence of the apparent Knight shift, $\Delta f/f_{o}$ in different magnetic fields $B_{ext}$, and utilize the fact that $\Delta f/f_{o}= K +\Delta \nu_{Q}^{(2)}/B_{ext} \rightarrow K$ in the limit of large $B_{ext}$.\cite{takigawa}  Since $\Delta \nu_{Q}^{(2)} \propto 1/B_{ext}$, a plot of $\Delta f/f_{o}$ as a function of $1/B_{ext}^{2}$ becomes a straight line, as shown in Fig.6.  By extrapolating the linear fit to the large field limit $1/B_{ext}^{2}\rightarrow 0$, we obtain $K = 0.14\%$ at 77~K.  In practice, since $^{75}K_{ab}$ is very small in the present case, we needed to carry out FFT measurements of the central peak shift $^{75}\Delta f^{center}$ to achieve high precision, as shown in the inset of Fig.6.

The advantage of this approach is that we don't need to know the details of the polar angle dependence of the EFG tensor; without making any assumptions on $\Delta\nu_{Q}^{(2)}$, one can experimentally deduce $^{75}K_{ab}$.  Alternatively, one can try to simulate the whole central transition\cite{grafe} by choosing both the  $\Delta\nu_{Q}^{(2)}$ and the Knight shift tensors as free parameters in the entire polar angle phase space.  However, this approach is known to be less accurate unless the Knight shift tensor is isotropic and the nuclear quadrupole interaction has very little distribution in its magnitude.  Neither of these conditions are satisfied in the present case.

The temperature dependence of $^{75}K_{ab}$ in the aligned polycrystalline sample is compared with that of $^{19}K_{ab}$ in Fig.3.  The agreement in temperature dependence is satisfactory.  A bonus of  $^{75}K_{ab}$ measurements is that, since the hyperfine coupling is already estimated as $^{75}A_{hf,ab} \sim 2.6$~Tesla/$\mu_{B}$,\cite{bafeas} we can estiamate the magnitude of spin susceptibility $\chi_{spin}$ using Eq.(3).  The conversion of $^{75}K_{ab}$ to $\chi_{spin}$ is shown in Fig.3: $\Delta ^{75}K_{ab} = 0.048$~\% translates to $\Delta \chi_{spin} = 1 \times 10^{-4}$~emu/mol-Fe.  At room temperature, we estimate the magnitude of spin susceptibility  $\chi_{spin} \sim 1.8 \times 10^{-4}$~emu/mol-Fe for LaFeAsO$_{0.9}$F$_{0.1}$.  This magnitude is about factor two larger than high $T_c$ cuprate superconductor, $\chi_{spin} \sim 1 \times 10^{-4}$~emu/mol-Cu.\cite{mila}

\section{Summary}
It is usually very difficult to nail down the intrinsic temperature dependence of $\chi_{spin}$ convincingly in correlated electron superconductors based on SQUID measurements alone.  This is in part because the large effects of superconducting diamagnetism below $T_c$ prevent us from estimating the Curie contributions from defects and impurities.  The advantage of the NMR approach is that nuclear spins can probe the local electronic properties without these extrinsic contributions.  From the $^{19}$F and $^{75}$As NMR measurements, we showed that $\chi_{spin}$ in LaFeAsO$_{0.9}$F$_{0.1}$ decreases with temperature, and levels off below $\sim 50$~K.  We also estimated the pseudo-gap $\Delta_{PG}/k_{B} \sim 140$~K.  At room temperature, $\chi_{spin} \sim 1.8 \times 10^{-4}$~emu/mol-Fe is comparable in magnitude as that of high $T_c$ cuprates.  We note that BaFe$_{1.8}$Co$_{0.2}$As$_2$ shows qualitatively the same behavior with much larger pseudo-gap $\Delta_{PG}/k_{B} \sim 560$~K, and the electronic crossover to a $\chi_{spin}=constant$ regime with decreasing temperature appears to be a generic property shared by various iron-pnictide superconductors.  Instead of concluding, we point out two major open questions.  First, what is the mechanism of the pseudo-gap phenomenon?  Second, what is the nature of the electronic state in the low temperature regime $\chi_{spin}=constant$? 

\section*{Acknowledgment}
T.I. acknowledges financial support from NSERC, CFI and CIFAR.  Research sponsored by the Division of Materials Science and Engineering, Office of Basic Sciences, Oak Ridge National Laboratory is managed by UT-Battelle, LLC, for the U.S. Department of Energy under contract No. DE-AC-05-00OR22725.  A portion of this work was performed by Eugene P. Wigner Fellows at ORNL.\\


\section*{Figure Captions}

Fig.1\\
 The crystal structure of LaFeAsO$_{0.9}$F$_{0.1}$. Fe atoms form a square-lattice within the ab-plane.  F atoms are located directly above/below Fe, while As atoms are located above/below the center of a square  formed by Fe.\\

Fig.2\\
$^{19}$F NMR lineshapes of ab-plane aligned polycrystalline sample of LaFeAsO$_{0.9}$F$_{0.1}$ at 290~K (top), 185~K, 80~K, 50~K, and 30~K (bottom).  All lineshapes were obtained by FFT of spin echo signals in $B_{ext}\sim 2.408$~Tesla.  For clarity, results above 30K are shifted vertically.  Gray dashed arrow and red arrow mark the unshifted frequency $^{19}f_{o} = 0$ where $^{19}K_{ab} = 0$, and the peak frequency $^{19}f$ at 30~K, respectively.\\

Fig.3\\
Left axis: $^{19}$F NMR Knight shift $^{19}K_{ab}$ (red bullets) measured for ab-plane aligned polycrystalline sample of LaFeAsO$_{0.9}$F$_{0.1}$ ($T_{c}\sim 27$~K).  Right axis: $^{75}$As NMR Knight shift $^{75}K_{ab}$ measured for the same sample (blue open squares).  The dashed line shows the contribution of the chemical shift $^{19}K_{chem} \sim 0.045$~\%, or equivalently, $^{75}K_{chem} \lesssim 0.1$~\%.   The red curve is the fit to Eq. (6) with $\Delta_{PG}/k_{B} = 140$~K.  Conversion to $\chi_{spin}$ is also shown (see section 4).\\

Fig.4\\
Resistivity measured for a sintered pellet of LaFeAsO$_{0.9}$F$_{0.1}$.  The sample is from the same batch of polycrystals used for NMR.  The inset shows the deviation from the $\rho \propto T^2$ law below $\sim50$~K.\\

Fig.5\\
(Color On-line) Main panel: Field swept $^{75}$As NMR lineshape of aligned polycrystals of LaFeAsO$_{0.9}$F$_{0.1}$ with $B_{ext}$// ab-plane.  The main peak at $\sim 7.5$~Tesla is from the $I_z = -\frac{1}{2}$ to $+\frac{1}{2}$ central transition.  The dashed grey arrow marks the central peak position in the absence of hyperfine interactions.   Inset: the same central transition in an enlarged scale (red).  The blue dashed line is the lineshape for an unaligned polycrystalline sample from the same batch. \\

Fig.6\\
(Color On-line) (a) FFT lineshape of the $^{75}$As central transition obtained at $B_{ext}\sim 7.45$~Tesla and $f = 54.763$~MHz.  (b) Apparent NMR Knight shift $^{75}\Delta f/^{75}f_{o}$ measured as a function of $1/(B_{ext})^{2}$.  Linear extrapolation to  $1/(B_{ext})^{2}=0$ gives the actual Knight shift $^{75}K_{ab}$.  Notice that the second order contribution of the nuclear quadrupole interaction, $\Delta \nu_{Q}^{(2)}$, is the dominant cause of the NMR frequency shift $^{75}\Delta f$ for typical magnetic field values of $B_{ext}\sim 7.45$~Tesla, and $^{75}K_{ab}$ is only a minor contribution.  The dominance of $\Delta \nu_{Q}^{(2)}$ makes accurate measurements of  $^{75}K_{ab}$ tricky.\\


\begin{thebibliography}{99}
\bibitem{imaiRunote} In Ref. [16], different temperature dependencies of $^{17}$O and $^{101}$Ru NMR Knight shifts data were interpreted throughly in the context of ferromagnetic correlations in all three bands, which turned out to be insufficient because quasi-1d electron-electron correlations effects exist in two of the bands due to Fermi-surface nesting effects.  However, all the conclusions are still valid as long as one reads off "ferromagnetic" correlations in the quasi-1d bands as "antiferromagnetic."
\bibitem{positiveAhf}There are two pieces of evidence that supports $A_{hf} >0$ at $^{19}$F sites.  First, a dynamical scaling law holds between $\chi_{spin}$ (i.e. $^{19}K$) and low frequency spin fluctuations $^{19}(1/T_{1}T)$.\cite{ahilan}  The latter has no ambiguities in the sign, and definitely decreases with temperature toward $T_c$.  As such $\chi_{spin}$ should also decrease toward $T_c$.\cite{ahilan}  Second, the sign of hyperfine coupling is known to be positive at $^{75}$As sites in FeAs layers.\cite{bafeas}  As shown in section 4, the $^{75}$As Knight shift shows identical behavior as $^{19}K$, and the former decreases toward $T_c$.
\bibitem{Kchem} From the dynamical scaling relation between $^{19}K$ and $^{19}(1/T_{1}T)$, we estimated $^{19}K_{chem}\sim 30$~ppm = 0.03\% for unaligned polycrystalline sample of LaFeAsO$_{0.89}$F$_{0.11}$.\cite{ahilan}  By repeating the same procedure for the present case, we obtain $^{19}K_{chem}\sim0.045 \pm 0.015$~\%.
\end{thebibliography}
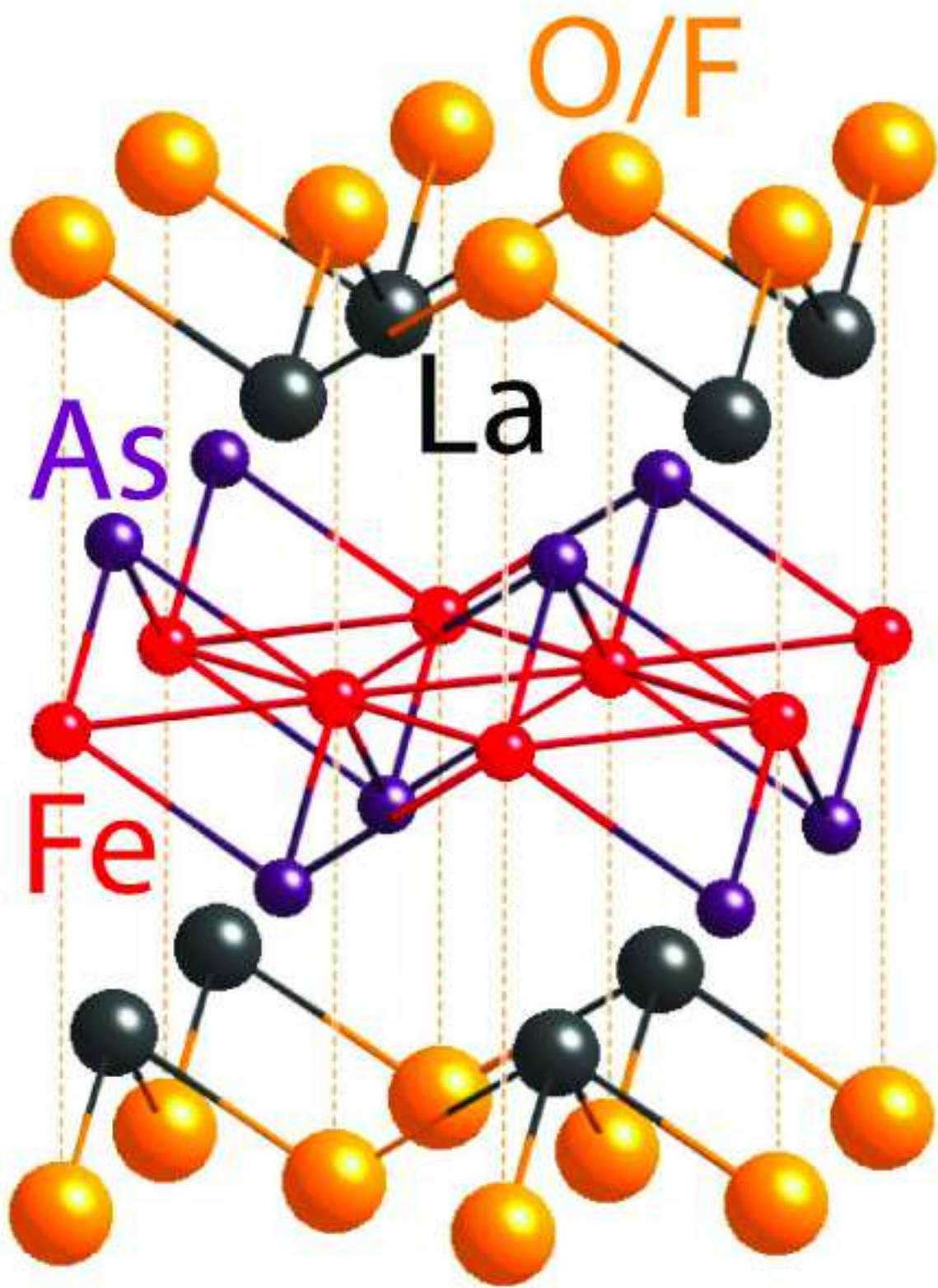
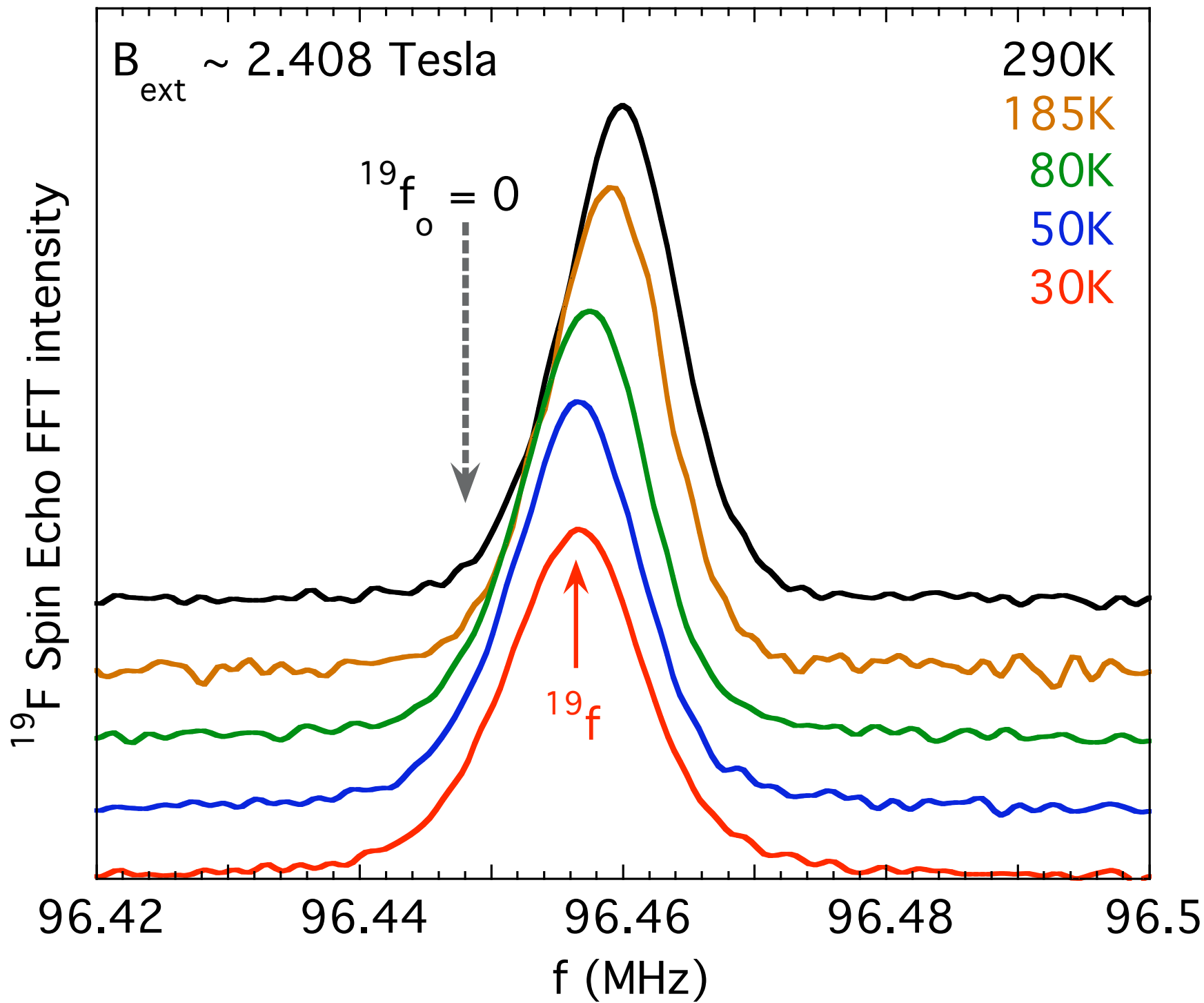
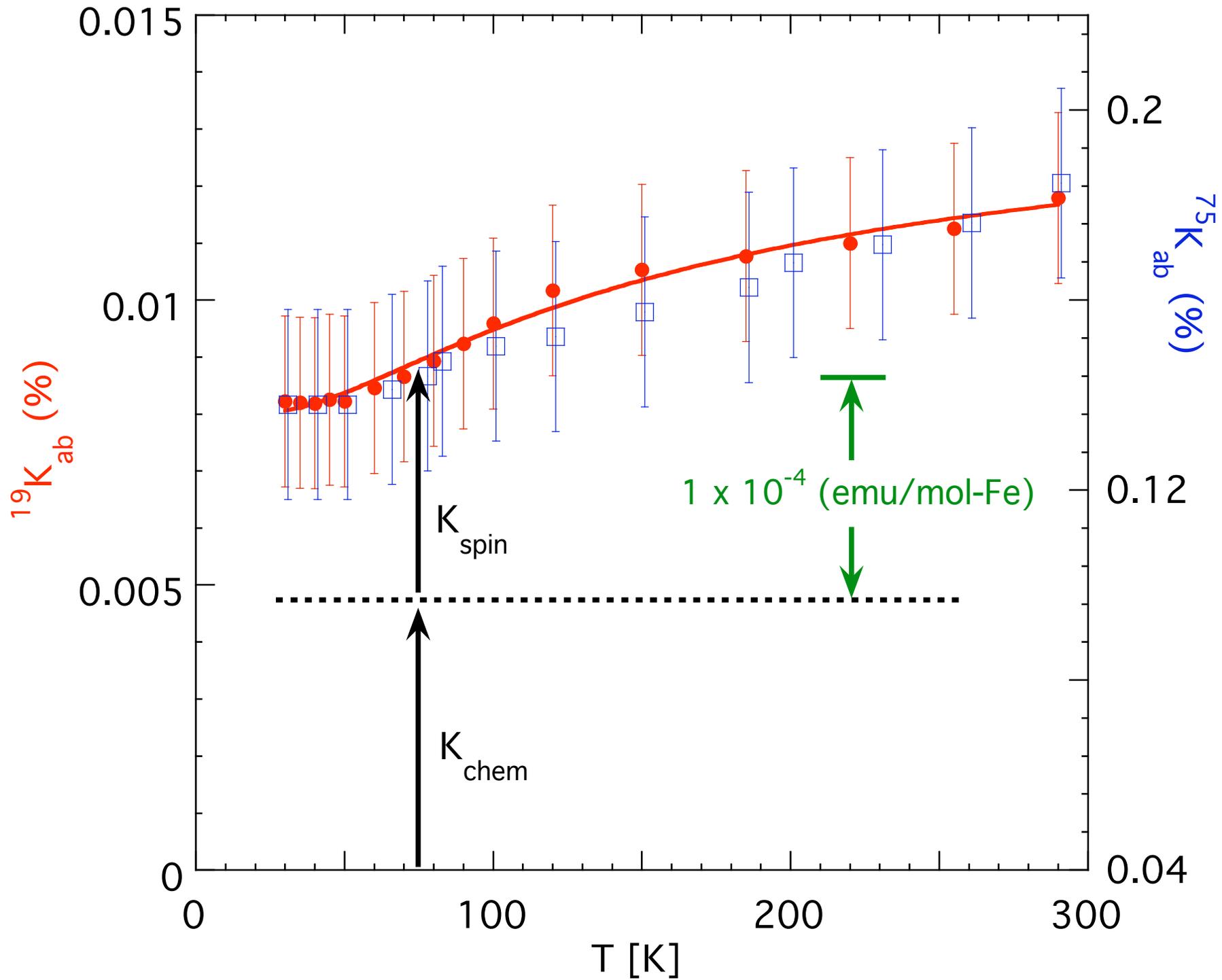
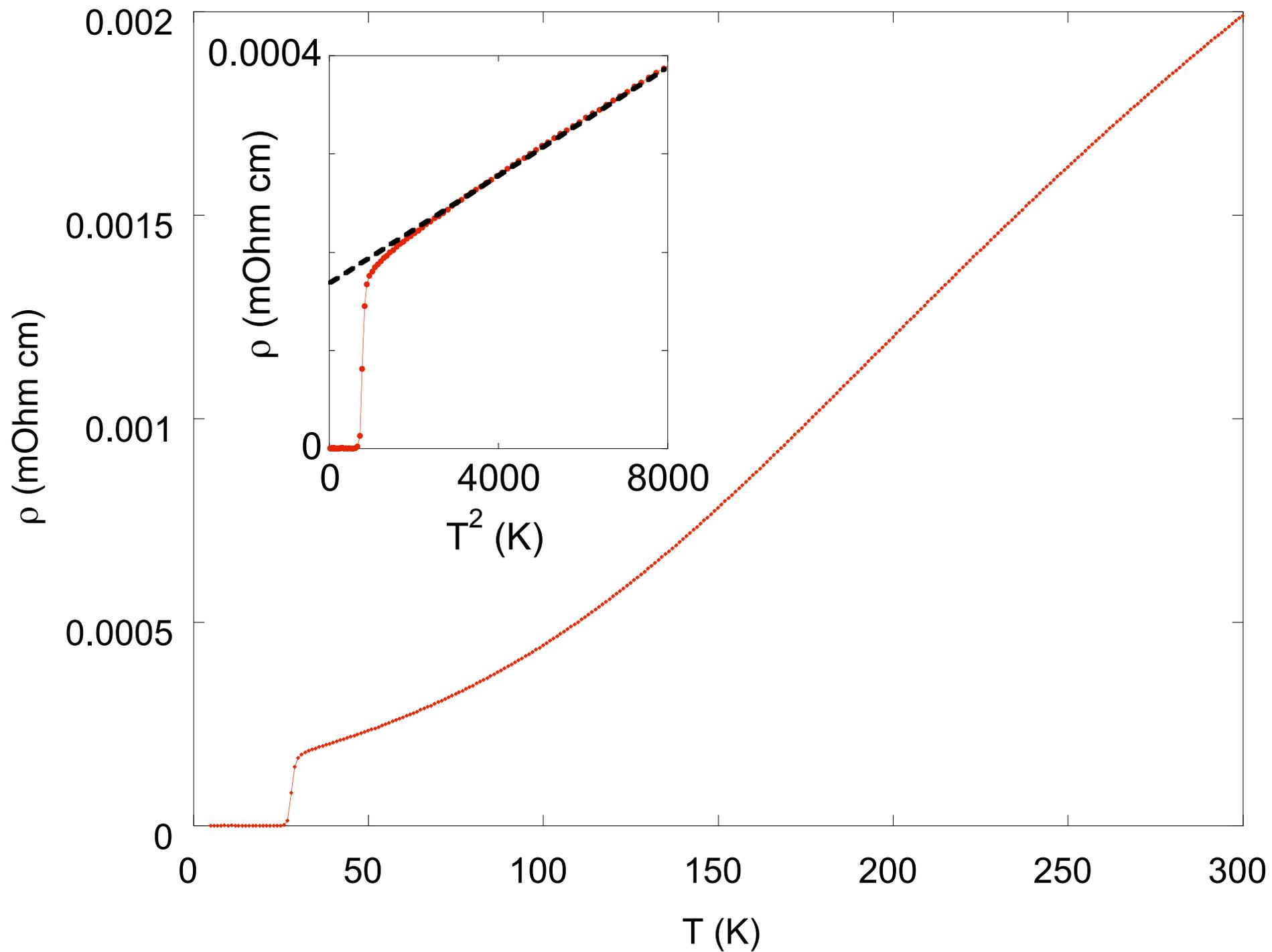
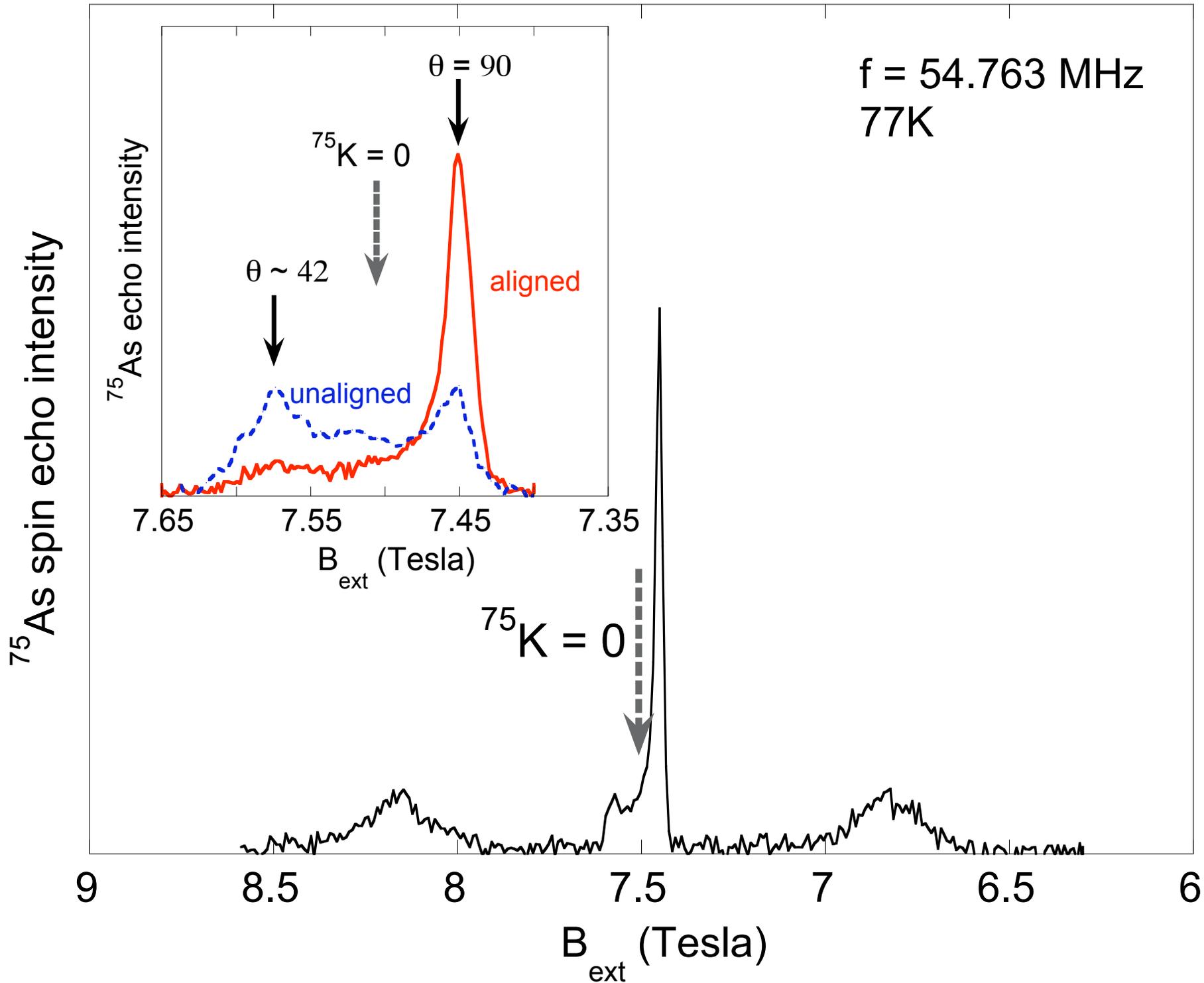
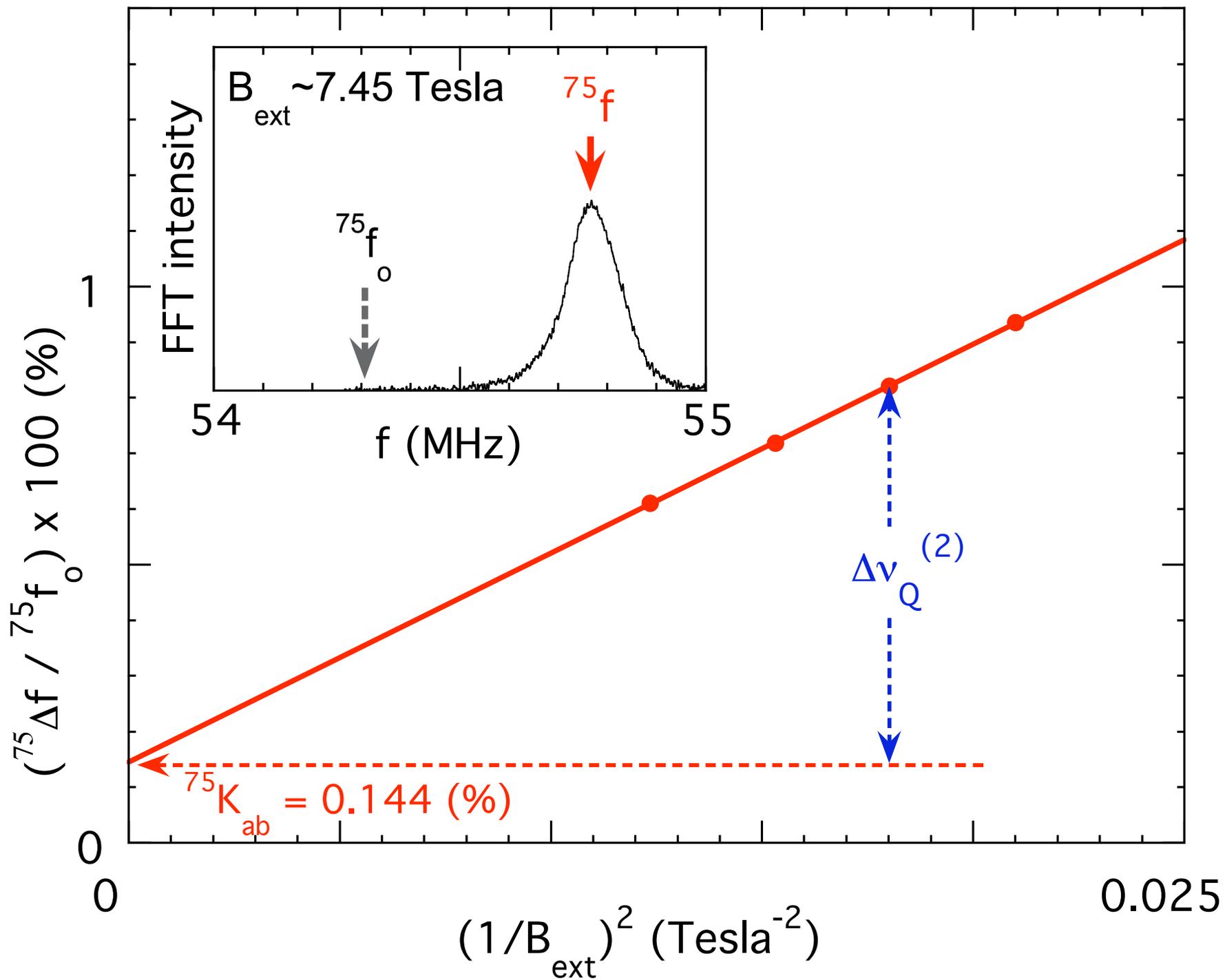
\end{document}